\documentclass[]{spie}  %>>> use for US letter paper

\usepackage{siunitx}
\usepackage{lineno}
\usepackage{amsmath,amsfonts,amssymb}
\usepackage{graphicx}
\usepackage{lipsum}
\usepackage{xspace}
\usepackage{sidecap}
\sidecaptionvpos{figure}{t}

\newcommand{\s}[1]{\S\ref{#1}}
\newcommand{\inch}{^{\prime\prime}}
\newcommand{\Vphi}{V-$\phi$\xspace}

\newcommand{\AdvACT}{AdvACT\xspace}

\title{Readout of two-kilopixel transition-edge sensor arrays for Advanced ACTPol}

\newcommand{\NSTRFGradStudents}{KTC, BJK, BLS, JTW, and SMS}

\author[a]{Shawn W. Henderson}
\author[a]{Jason R. Stevens}

\author[k]{Mandana Amiri}
\author[g]{Jason Austermann}
\author[g]{James A. Beall}
\author[e]{Saptarshi Chaudhuri}
\author[f]{Hsiao-Mei Cho}
\author[d]{Steve K. Choi}
\author[a]{Nicholas F. Cothard}
\author[d]{Kevin T. Crowley}
\author[g]{Shannon M. Duff}
\author[g]{Colin P. Fitzgerald}
\author[a]{Patricio A. Gallardo}
\author[k]{Mark Halpern}
\author[b,c]{Matthew Hasselfield}
\author[g]{Gene Hilton}
\author[d]{Shuay-Pwu Patty Ho}
\author[g]{Johannes Hubmayr}
\author[e,f]{Kent D. Irwin}
\author[a]{Brian J. Koopman}
\author[f]{Dale Li}
\author[d]{Yaqiong Li}
\author[j]{Jeff McMahon}
\author[h]{Federico Nati}
\author[a]{Michael D. Niemack}
\author[g]{Carl D. Reintsema}
\author[d]{Maria Salatino}
\author[i]{Alessandro Schillaci}
\author[h]{Benjamin L. Schmitt}
\author[d]{Sara M. Simon}
\author[d]{Suzanne T. Staggs}
\author[a]{Eve M. Vavagiakis}
\author[h]{Jonathan T. Ward}

\affil[a]{Department of Physics, Cornell University, Ithaca, NY, USA 14853}
\affil[b]{Department of Astronomy and Astrophysics, The Pennsylvania State University, University Park, PA 16802}
\affil[c]{Institute for Gravitation and the Cosmos, The Pennsylvania State University, University Park, PA 16802}
\affil[d]{Joseph Henry Laboratories of Physics, Jadwin Hall, Princeton University, Princeton, NJ, USA 08544}
\affil[e]{Department of Physics, Stanford University, Stanford, CA, USA 94305-4085}
\affil[f]{SLAC National Accelerator Laboratory, 2575 Sandy Hill Road, Menlo Park, CA 94025}
\affil[g]{NIST Quantum Devices Group, 325 Broadway Mailcode 817.03, Boulder, CO, USA 80305}
\affil[h]{Department of Physics and Astronomy, University of Pennsylvania, 209 South 33rd Street, Philadelphia, PA, USA 19104}
\affil[i]{Departamento de Astronom\'{i}a y Astrof\'{i}sica, Pontic\'{i}a Universidad Cat\'{o}lica, Casilla 306, Santiago 22, Chile}
\affil[j]{Department of Physics, University of Michigan, Ann Arbor, USA 48103}
\affil[k]{Department of Physics and Astronomy, University of British Columbia, Vancouver, BC, Canada V6T 1Z4}

\authorinfo{Further author information: (Send correspondence to S.W.H.)\\S.W.H.: E-mail: swh76@cornell.edu, Telephone: 1 607 255 0419}

\begin{document} 

\maketitle

\begin{abstract}
Advanced ACTPol is an instrument upgrade for the six-meter Atacama
Cosmology Telescope (ACT) designed to measure the cosmic microwave
background (CMB) temperature and polarization with arcminute-scale
angular resolution.  To achieve its science goals, Advanced ACTPol
utilizes a larger readout multiplexing factor than any previous CMB
experiment to measure detector arrays with approximately two thousand
transition-edge sensor (TES) bolometers in each 150 mm detector wafer.
We present the implementation and testing of the Advanced ACTPol
time-division multiplexing readout architecture with a 64-row
multiplexing factor. This includes testing of individual multichroic
detector pixels and superconducting quantum interference device
(SQUID) multiplexing chips as well as testing and optimizing of the
integrated readout electronics.  In particular, we describe the new
automated multiplexing SQUID tuning procedure developed to select and
optimize the thousands of SQUID parameters required to readout each
Advanced ACTPol array.  The multichroic detector pixels in each array
use separate channels for each polarization and each of the two
frequencies, such that four TESes must be read out per
pixel. Challenges addressed include doubling the number of detectors
per multiplexed readout channel compared to ACTPol and
optimizing the Nyquist inductance to minimize detector and
SQUID noise aliasing.
\end{abstract}


\section{Introduction}
\label{sec:intro}

Large arrays of low-temperature detectors are finding increasingly
wider use in the detection of radiation, from photons in the sub-mm
and gamma-rays, to neutrinos, and even dark matter.  In Cosmic
Microwave Background (CMB) and sub-mm astronomy, large arrays of
detectors 
permit substantial improvements in sensitivity and mapping speed if the 
individual detectors are limited by the photon noise background.  The most mature 
superconducting detector technology, the transition-edge sensor (TES)~\cite{Irwin95},
has demonstrated background limited performance across a range of bands and platforms.
The present generation of ground-based CMB
experiments (so-called Stage-III) plan to field roughly $10^{4}$
polarization sensitive detectors, and the CMB community is actively
planning a future Stage-IV effort to field $10^{5}$-$10^{6}$
detectors~\cite{Wu14}, many of which may populate the focal planes of a small
number of high-throughput telescopes~\cite{Niemack16}.

Present Stage-III ground-based efforts include Advanced ACTPol (\AdvACT) on the
six-meter Atacama Cosmology Telescope (ACT)~\cite{Henderson16},
BICEP3/Keck Array~\cite{Ogburn10,Staniszewski12},
CLASS~\cite{Essinger-Hileman14}, the Simons Array~\cite{Arnold14}, and
SPT-3G on the South Pole Telescope~\cite{Benson14} (balloon-borne experiments underway include
EBEX~\cite{Reichborn-Kjennerud10} and SPIDER~\cite{Filippini10}, and planned satellite missions include LiteBIRD~\cite{Matsumura14}).  All of these
experiments are fielding, or plan to field, large ($>10^{3}$
pixels/array) sub-Kelvin arrays of TES bolometers.  Broadly, these
experiments seek to map the microwave sky
($\sim$(30-300)~GHz) in both intensity and polarization on
arcminute scales to several degree angular scales to improve our
understanding of cosmology, the evolution of structure in the
universe, galaxy clusters, and millimeter sources.  High
signal-to-noise polarization maps over degree angular scales in
particular have the potential to provide unique sensitivity to the
signatures of gravitational waves produced in the very early 
Universe, such as those produced in some inflationary models~\cite{Abazajian15}.

The readout of these large sub-Kelvin detector arrays is complex,
requiring novel superconducting electronics with thousands of
components and interconnects.  To reduce the cost and complexity of the readout and the
thermal load on the cryogenics presented by warm electronics and
cables, the detectors on these arrays must be multiplexed, with many
groups of detectors read out and controlled by a much smaller number of
wires and devices.  In this proceedings, we describe the
implementation and testing of the custom readout developed for
the first \AdvACT high frequency (HF) array, which consists of
$2024$ TESes operated at $\sim100$~mK.

In \s{sec:advact} we describe the \AdvACT experiment.  In
\s{sec:multiplexing} we discuss multiplexing generally and then the
specific multiplexing implementation chosen for \AdvACT in
\s{sec:implementation}.  In \s{sec:characterization}, we describe
screening and characterization performed on readout components for the
first \AdvACT array, and we conclude in \s{sec:summary} with a
discussion of the implications of this work for future efforts.

\section{Advanced ACTPol}
\label{sec:advact}

\AdvACT is a receiver upgrade for the Atacama Cosmology Telescope
(ACT) which builds on the success of the Atacama Cosmology Telescope
Polarization-sensitive receiver (ACTPol)~\cite{Niemack10}.  ACTPol
observed with three arrays of TES bolometers in two frequency bands
centered at $90$ and $150$~GHz from 2013 until 2016~\cite{Naess14} from the Atacama
desert in Northern Chile.  
\AdvACT aims to map
roughly half of the microwave sky from the ground in five frequency
bands with four upgraded arrays of TES bolometers~\cite{Henderson16}.  A
high frequency (HF) array will observe simultaneously at
$150$/$230$~GHz, two medium frequency (MF) arrays will observe
simultaneously at $90$/$150$~GHz, and one low frequency (LF) array
will observe at $28$/$41$~GHz.  The \AdvACT arrays will be deployed as a staged
upgrade of the existing ACTPol receiver, and the first deployed array
will be the HF, which will replace the first ACTPol array in mid-2016.

Each array is fabricated on a single monolithic $150$~mm
wafer~\cite{Duff16}, enabling much denser pixel packing than
achievable in the ACTPol arrays which were composed of tiles
fabricated on $3\inch$ wafers.  The higher pixel density of the
\AdvACT arrays, coupled with the fact that multichroic arrays have
twice as many detectors per pixel, nearly doubles the number of
detectors in the HF and MF arrays relative to the ACTPol arrays.  The
ACTPol readout was limited to a maximum of $1024$ channels per array,
which is far short of the $2024$ channels required for the \AdvACT HF array.
This limitation necessitated a new cryogenic multiplexing architecture
(Sec~\s{sec:multiplexing}), as well as new cryogenic interfaces,
electronics, and software (Sec~\s{sec:implementation}).

\section{Multiplexing Overview}
\label{sec:multiplexing}

Multiplexing, which enables the readout of a larger number of signals
than the number of signaling lines, is a necessary step in the
readout of large arrays of cryogenic detectors.  Several schemes exist
for multiplexing the signals from large arrays of TES bolometers.  At
present, the most mature techniques employ superconducting quantum
interference device (SQUID) amplifiers arranged in Frequency- or
Time-Division Multiplexing (FDM~\cite{Dobbs12} and
TDM~\cite{deKorte03}) architectures.  In TDM, each TES bolometer is
sampled sequentially at MHz frequencies, while in FDM, the signal from
each TES bolometer is modulated with a carrier wave at a unique
frequency (in the MHz) and then demodulated in warm electronics to
extract each detector's individual signal.  The multiplexing factor
(MUX factor) is defined as the number of detectors per readout
channel.  MUX factors as high as 40 have been achieved on fielded TES
arrays by SCUBA2~\cite{Holland13} using TDM and as high as 16 by
EBEX~\cite{MacDermid14} using MHz FDM.  Next generation readout
systems are in development for CMB experiments utilizing MHz FDM that will operate with MUX
factors as high as 64~\cite{Bender14}, as well as TDM systems with MUX factors
as high as 128~\cite{Prêle2016}.

\begin{SCfigure}
\centering
\includegraphics[width=0.55\linewidth,keepaspectratio]{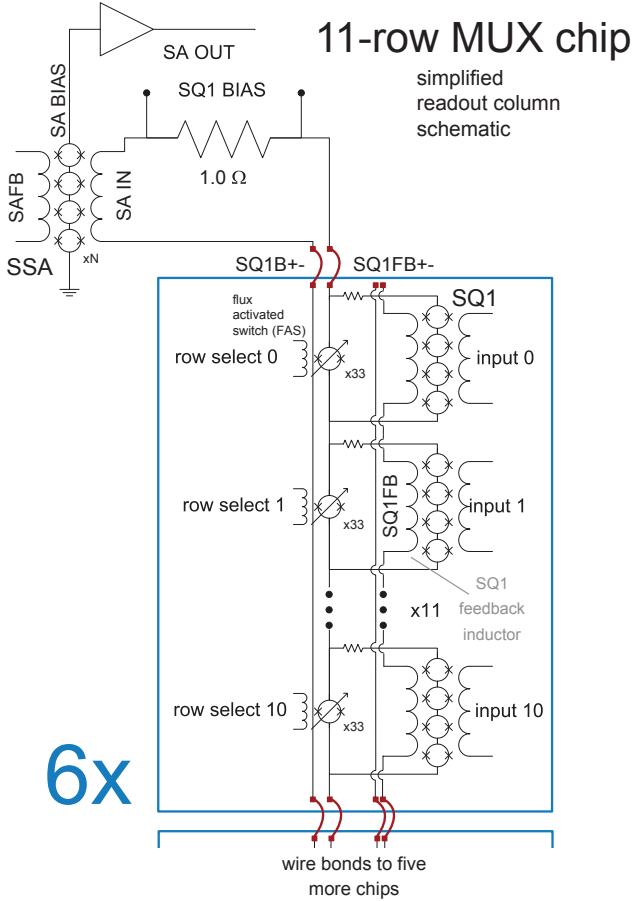}
\caption{
  Schematic of the SQUID-based multiplexing circuit used to read out
  one of the 32 columns of 64 TESes in the \AdvACT HF array~\cite{Doriese15}.  Each TES is
  inductively coupled to its own first stage SQUID series array (SQ1).
  Each SQ1 is shunted by a flux activated switch (FAS).  Each of the 64 SQ1/FAS
  pairs are connected in series to form the 64 rows.  Each row is
  sequentially addressed by driving its FAS with the flux required to
  drive it normal, resulting in an approximate voltage bias of its
  companion SQ1 (when normal, $R_{\rm FAS}/R_{\rm dyn}^{\rm SQ1}\sim(3-7)$~\cite{Doriese15}).  Channels not being read out are decoupled from the
  circuit by applying a flux for which their FASes remain
  superconducting, shorting the inputs of their SQ1s.  Each column's
  SQ1/FAS chain is biased in parallel with a \SI{1}{\ohm} bias
  resistor and read out through a $\sim$\SI{1}{\kelvin} SQUID series
  array (SSA) by the warm preamplifier chain of the Multi-Channel
  Electronics (MCE)~\cite{Battistelli08,Battistelli08_2}.  The TESes are read out by the MCE in a
  flux-locked loop (FLL) which servoes the output of the SSA by means
  of the SQ1 feedback to maintain linearity in the SQUID readout
  chain.  Each readout column is implemented with six 11-row
  multiplexing (MUX) chips, only the first of which is shown here.  On
  each column six MUX chips are connected in series using
  superconducting aluminum wirebonds for 66 rows in total, although only
  64 rows are needed to fully channelize the \AdvACT HF array.  The
  SQ1 bias and feedback circuits are closed by means of
  superconducting Al shorting wirebonds at the end of
  each column (not shown).
}
\label{fig:muxschematic}
\end{SCfigure}

\subsection{SQUID Electronics}

The \AdvACT arrays will be multiplexed using a new low-frequency
TDM architecture with two stages of SQUID amplification.
Figure~\ref{fig:muxschematic} shows a schematic of the multiplexing
circuit (MUX).
Each TES in the array is inductively coupled to its own first-stage DC
SQUID series array (SQ1). The series array is connected in parallel
with a Josephson junction array in the form of a Zappe interferometer~\cite{Zappe77}
which acts as a flux-activated switch (FAS).
Each SQ1/FAS pair forms a channel of the readout, with many
channels connected in series to form ``columns'' of channels.  
A bias current is applied to the chain of SQ1s/FAS units on each
column in parallel with a \SI{1}{\ohm} resistor. The dynamic
resistance of the SQ1/FAS chain is large compared to \SI{1}{\ohm}, so it is
approximately voltage biased. All but one FAS are left in the
superconducting state, so that the current does not pass through their
companion SQ1s. One FAS associated with the addressed channel is
flux-biased in its normal state, so that the full voltage is dropped
across that SQ1/FAS unit.
Each column of
the multiplexer is read out by a unique SQUID series array at
$\sim$\SI{1}{\kelvin}, which provides a second stage of amplification before the
warm electronics.  Channels are switched across columns by connecting
the addressing coil of each FAS in series with one other FAS in every
column, forming ``rows'' of channels.

The \AdvACT HF MUX is implemented on $32$ columns and $64$ rows using
this architecture.  Variants of the architecture have been used
successfully, albeit with lower MUX factors, to read out arrays of
X-ray TESes~\cite{Doriese15} as well as a 1280-pixel array of
TESes observing the CMB at \SI{95}{\giga\hertz} in
BICEP3~\cite{Ahmed14}.  Early implementations have shown significant
advantages over the three-stage SQUID architecture used in ACTPol,
including higher bandwidth, lower power dissipation, and a reduction
in total readout noise for comparable MUX factors~\cite{Doriese15}.

\section{Advanced ACTPol Readout Implementation}
\label{sec:implementation}

In order to be read out, each TES in the HF array must be electrically
connected to the MUX, and then to the warm readout electronics.
\s{subsec:coldelectronicsandinterfaces} and \s{subsec:nyquist} below
describe the custom interfaces developed to interface each \AdvACT array
with the warm electronics, while \s{subsec:mce} and \s{subsec:tuning} 
describe the warm electronics and their operation.

\subsection{Cold Electronics and Interfaces}
\label{subsec:coldelectronicsandinterfaces}
\begin{figure}[ht]
   \begin{center}
   \begin{tabular}{c}
     {\includegraphics[width=16cm]{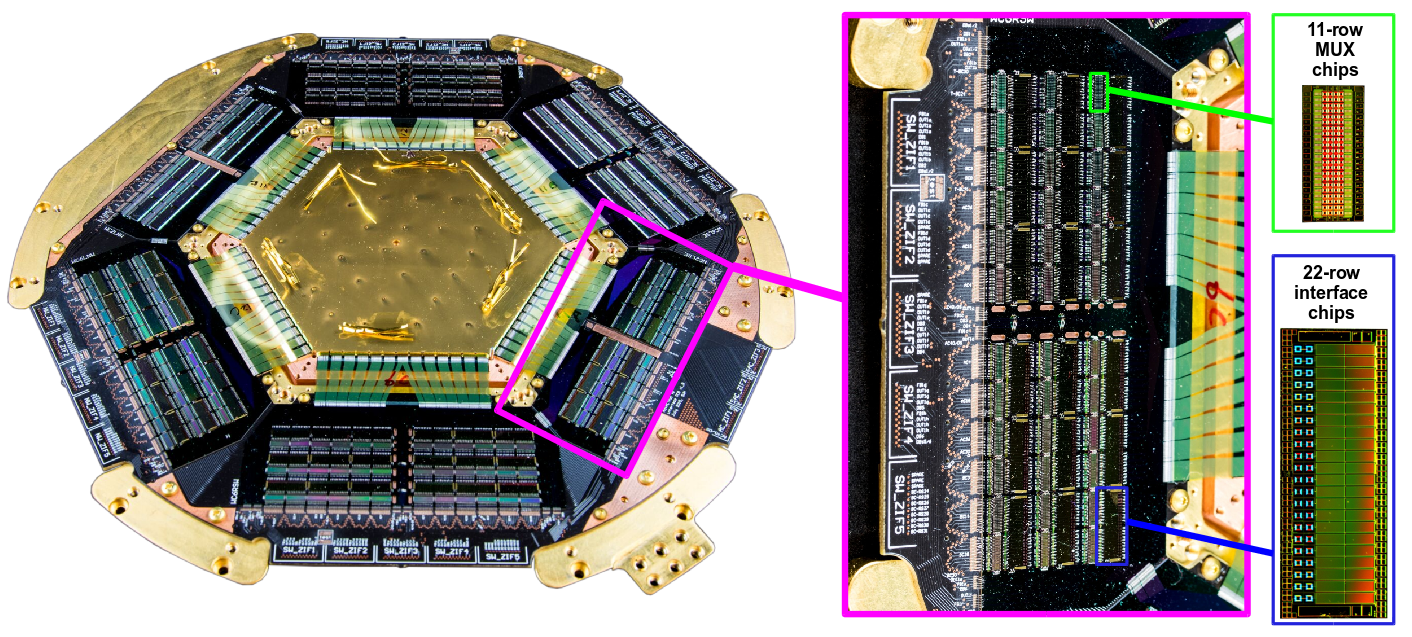}}
   \end{tabular}
   \end{center}
   \caption[example] 
   { \label{fig:hfphotos} 
     \emph{Left:} The fully integrated \AdvACT HF focal plane,
     pre-deployment.  Viewed from the back side of the array package.
     The hexagonal array, at center, is connected to eight large
     silicon wiring chips through superconducting flex cables.  The
     wiring chips are glued to the surface of a black ring-shaped PCB
     mounted to a gold-plated Cu support ring, which interfaces the
     array and cold electronics with ACT's dilution refrigerator's
     \SI{100}{\milli\kelvin} cold-plate.  \emph{Right:} A closeup of
     one of the largest silicon wiring chips used to read out the
     \AdvACT HF array, implementing six columns of the 64-row readout.
     Closeups are also shown of the MUX and detector bias interface
     chips used to read out the array TESes which are affixed directly
     to the surface of the wiring chips.  Each readout column requires
     six 11-row MUX chips and three 22-row interface chips.
   }
\end{figure}

While \AdvACT reuses several of the cryogenic interfaces and
PCBs from the ACTPol arrays, the new MUX architecture and \SI{150}{\milli\metre}
wafer format necessitated a total redesign of the \SI{100}{\milli\kelvin} PCBs and
interfaces for the HF array.  Figure~\ref{fig:hfphotos} shows the
focal plane of the integrated HF array.  In contrast to ACTPol's
folded vertical readout design~\cite{Grace14}, the entirety of the new \AdvACT
\SI{100}{\milli\kelvin} readout is planar.

Each of the $503$ feedhorn-coupled pixels in the HF array contains four TESes, with
two coupled to orthogonal polarization signals in the \SI{150}{\giga\hertz} 
band, and the other two coupled to orthogonal polarization 
signals in the \SI{230}{\giga\hertz} band~\cite{Datta14}.  Three additional pixels at mid-radius
in the HF array are not coupled to feedhorns, allowing for dark measurements on those pixels' 12 TESes and mechanical support.
Signals from the TESes are connected
from each pixel to pairs of bondpads on the edges of the array using 
between $0.2$ and \SI{6.3}{\centi\metre} of \SI{5}{\micro\metre} wide niobium microstrip lines.
This results in $338$ pairs of bondpads on each side of the array. 
The pads are \SI{120}{\micro\metre} wide and \SI{400}{\micro\metre} long and arranged
into two rows, with an adjacent pad pitch of \SI{140}{\micro\metre}.  Each side 
of the array receives only the signals from one optical band, and 
orthogonal polarizations within each band for the same pixel are 
routed together on adjacent microstrip lines to adjacent pad pairs.

From the pads on the sides of the array, the TES signals are connected
into custom superconducting flexible cables (flex) developed
originally for ACTPol~\cite{Pappas15} with superconducting aluminum
wirebonds.  Each flex cable is \SI{2}{\centi\metre} long and routes
the TES signals through \SI{50}{\micro\metre} wide superconducting
aluminum traces on a \SI{70}{\micro\metre} pitch over a flexible
polyimide substrate.  Each flex cable has thick integrated
\SI{5}{\milli\metre} long silicon stiffeners at either end.  One
stiffener is rubber-cemented to the silicon ``wings'' attached to the
HF array~\cite{Ward16,Li16}, while the other is rubber cemented to a large
readout PCB with a hexagonal inner cut-out.  Each side of the flex contains
wirebond pads with the same dimensions and pitch as the pads on the array.

From the flex, the TES signals are aluminum wirebonded to a matching
set of niobium pads on large silicon wiring chips that are rubber
cemented to the PCB and abut the ends of the flex.  The eight wiring
chips on the PCB that interface with the flex further route the TES
signals to bond pad pairs aligned with input pads on the sides of
the MUX and interface chips (see Figure~\ref{fig:hfphotos}).  Signals are routed on the wiring chips 
through $0.9$ to \SI{5}{\centi\metre} long pairs of \SI{15}{\micro\metre} wide niobium
lines on a \SI{30}{\micro\metre} pitch.  Four large
(\SI{10.5}{\centi\metre} x \SI{4}{\centi\metre}) wiring chips are used
to route signals to six columns of the MUX apiece, while four smaller
(\SI{4.9}{\centi\metre} x \SI{3}{\centi\metre}) wiring chips are each
used to route signals to two columns of the readout.  In order to pack
the readout as densely as possible, TES signals are interleaved
between 
MUX and interface chips
on each wiring chip, with orthogonal
polarization pairs from the same pixel always routed side-by-side into
the same readout column together.  In this way, detectors in the same
pixel observing orthogonal polarizations in the same optical band are
read out through the same amplifier chain, on neighboring rows.

The MUX and interface chips fabricated at NIST/Boulder
are epoxied to the wiring chips and implement the first stage of the
MUX.  TES signals are first wirebonded from the wiring chips into a
22-channel bias interface chip (\SI{5}{\milli\metre} x
\SI{13.45}{\milli\metre}), which connects the TES in parallel with a
cryogenic shunt resistor ($\sim$\SI{0.2}{\milli\ohm}) and in series with an optional
bandwidth-limiting ``Nyquist'' inductor~\cite{Hilton2005}.  The TESes are then each connected via
a pair of wirebonds from each channel of the interface chips to a channel
on an 11-channel MUX chip (\SI{3}{\milli\metre} x
\SI{6.6}{\milli\metre}, see Figure~\ref{fig:muxschematic}) containing the SQ1 and FAS for each TES.  The
NIST interface chips and MUX chips (NIST mask name ``mux15b'') are
designed to be connected in series, which for the HF readout requires three
interface chips and six MUX chips per readout column.  
One of the extra rows
is connected to a row of ``dark SQUID'' (DSQ) channels threading the
MUX that are not connected to TESes and can be optionally read out,
while the other spare row is left unconnected.  In addition to this optional DSQ row, there
are 24 individual channels without detectors scattered throughout the
MUX that are also designated DSQ channels and will be read out with the
array.

Along the outer perimeter of the PCB, on the edges of the wiring chips,
additional aluminum wirebonds distribute SQ1 bias and feedback lines,
TES bias lines, and the 64 FAS addressing lines from the PCB into the
appropriate MUX and interface chips through additional niobium wiring on
the wiring chips.  There is one independent SQ1 feedback and SQ1 bias line
for every one of the 32 readout columns, but only 24 TES bias lines
for the entire array.  Although each column in the readout does not
have a dedicated TES bias line, each bias line is connected only to
TESes in the same optical band, motivated by the significantly
different bias conditions expected due to predicted differences in
atmospheric loading at 150 and \SI{230}{\giga\hertz}.  Five adaptor PCBs
connected to the large array PCB through copper flexible cables with
zero insertion force (ZIF) connectors distribute the SQUID and TES
control lines into PCBs anchored at \SI{1}{\kelvin} through NbTi
cables.

The PCBs at \SI{1}{\kelvin} contain the final stage of SQUID
amplification and further route the amplified signals, as well as the
other SQUID and TES control lines, through superconducting NbTi and
low thermal conductivity manganin twisted-pair cables to room
temperature electronics which control the multiplexing~\cite{Battistelli08,Battistelli08_2}.  The
final \SI{1}{\kelvin} SQUID stage is implemented using compatible
NIST/Boulder SQUID Series Array Modules (SSAMs)~\cite{Doriese15}.  Each
SSAM contains eight SQUID series arrays, each of which amplifies a unique
readout column in the HF array.  The \SI{1}{\kelvin} \AdvACT PCBs are
functionally identical to those used in ACTPol, except for the new SQUID series
arrays, which have been tailored to work with the
\SI{100}{\milli\kelvin} MUX chips in the new two-stage TDM
architecture, and different resistances for cryogenic bias resistors
on the SQ1 bias lines.

\subsection{Nyquist inductance optimization}
\label{subsec:nyquist}

The signal from each TES in the array is routed through an
interface chip before being inductively coupled into a SQ1.  Which
pads are bonded on the interface chip 
for a channel determine how much extra
inductance is added in series with each array TES.  The purpose of
this optional extra inductance is to bandwidth limit the signals from
the TES above the Nyquist frequency of the MUX.  The planned HF
multiplexing rate is \SI{7.8}{\kilo\hertz}, although timestreams will
be further downsampled to $\sim$\SI{400}{\hertz}.  While the TESes
exhibit non-negligible current noise at \SI{10}{\kilo\hertz}, the TES
bandwidth cannot be arbitrarily restricted with a series Nyquist
inductance without jeopardizing the stability of the device,
particularly when operated low on the transition~\cite{Irwin05} (
$\lesssim40$\% the normal TES resistance).  To allow for the optimization
of the Nyquist inductance for each array, the \AdvACT interface chips
were designed with three wirebond-selectable Nyquist inductance
options realized by N-turn spiral inductors : 0-turn (no added
inductance), 9-turn ($\sim$\SI{60}{\nano\henry} of added inductance)
and 17-turn ($\sim$\SI{200}{\nano\henry} of added inductance).

\begin{figure} [ht]
\centering
\begin{tabular}{@{}c@{}}
\includegraphics[width=8cm]{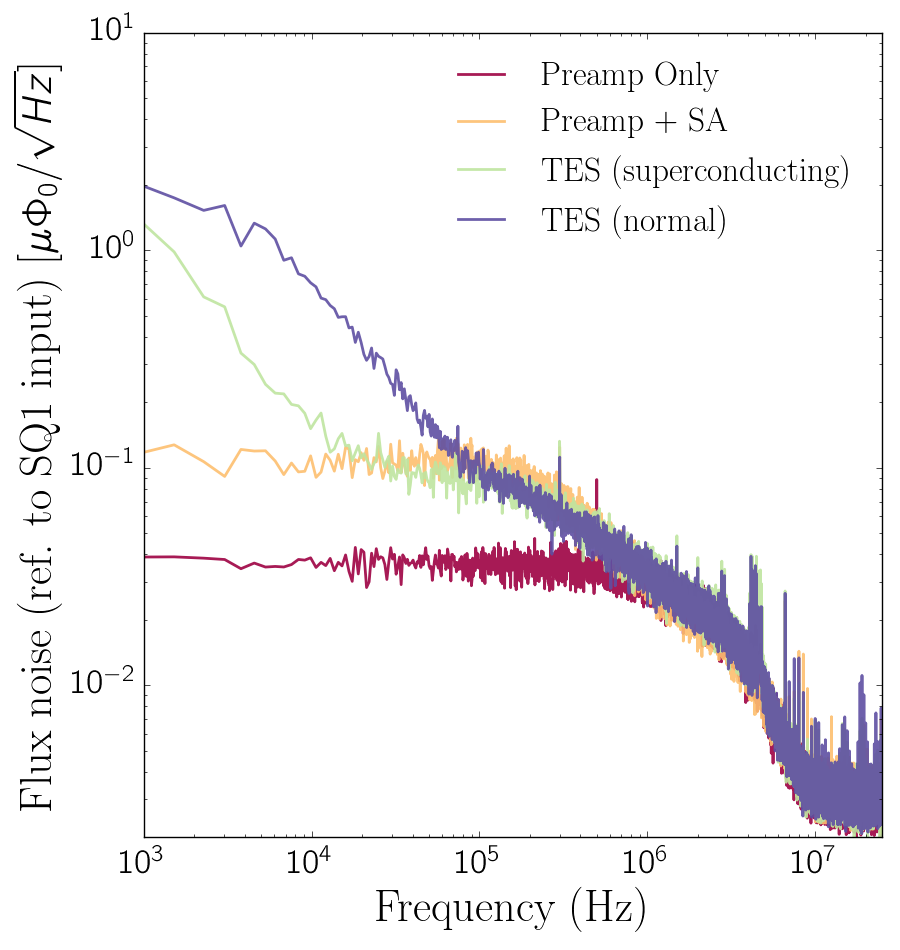}
\end{tabular}
\begin{tabular}{@{}c@{}}
\includegraphics[width=8cm]{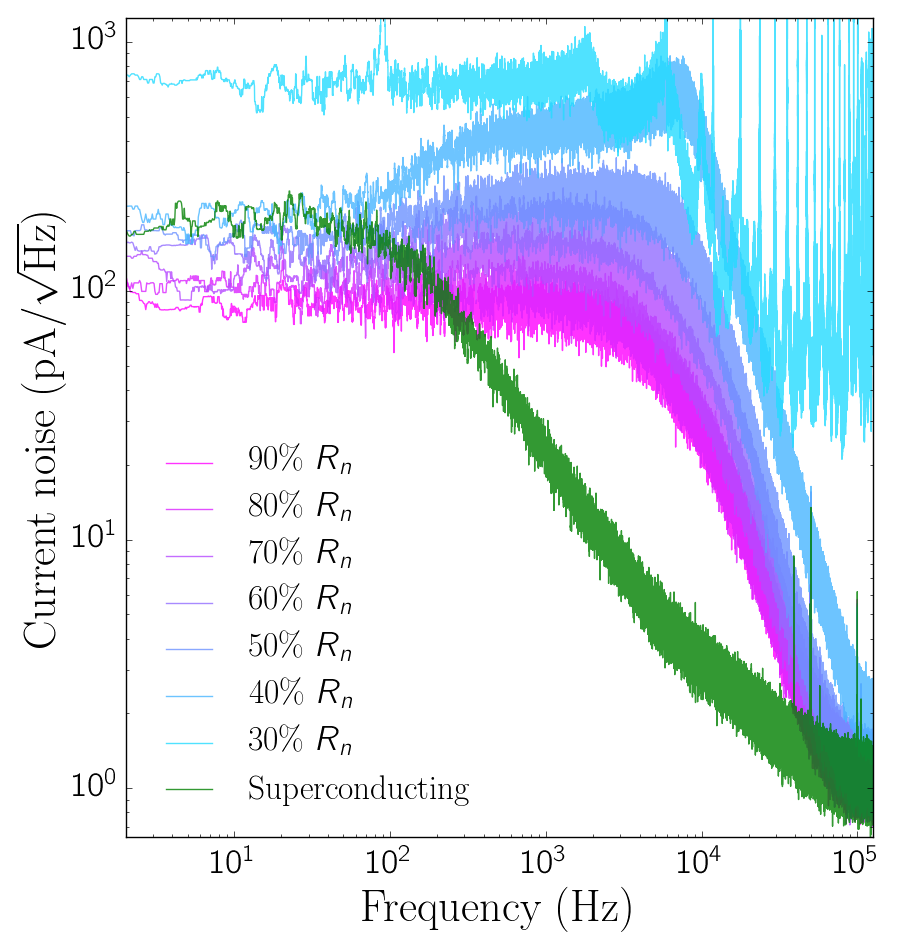}
\end{tabular}
\caption{
\emph{Left:}.  Open loop noise measurements taken on a testbed built
to match the fielded AdvACT readout illustrating the bandwidth of
successive stages of the multiplexer, including the warm MCE
preamplifier chain (Preamp).  Data were sampled at
\SI{50}{\mega\hertz}, and the noise amplitude has been referred to
flux at the input coil of the first-stage SQUID.  
The analog open-loop bandwidth of the multiplexer limits 
the multiplexing rate to \SI{7.8}{\kilo\hertz} for the \AdvACT 
HF array, despite the much higher demonstrated bandwidth
of the MUX architecture~\cite{Doriese15}.
At these very high sampling rates, the flux noise due to the superconducting/normal
TES is heavily suppressed by the TES bias circuit.
\emph{Right:}
Current noise measurements taken through the superconducting
transition for a prototype HF \SI{230}{\giga\hertz} AlMn TES
fabricated on a single pixel test die~\cite{DLi16} connected to the MUX through a
9-turn Nyquist inductor and operating at a bias power corresponding to
the best expected observing conditions in the field, sampling at
\SI{250}{\kilo\hertz}.  These measurements and similar measurements
performed on \SI{150}{\giga\hertz} HF devices were used to inform the
final TES designs as well as the selected Nyquist inductances for all
HF channels.  This particular prototype device exhibits detector
instability low on the transition ($R_{\rm TES}$ $\lesssim40$\% $R_{n}$).}\label{fig:MUXandTESnoise}
\end{figure}

Single pixel noise measurements indicate a modest reduction
in measured noise (4-10\%) for devices connected in series with a 9-turn
inductor versus no inductance when multiplexed at
\SI{7.8}{\kilo\hertz}.  Conversely, a visual examination of time
streams multiplexed at \SI{250}{\kilo\hertz} indicated the onset of
instability higher on the transition for identical devices 
connected in series with the 17-turn inductor, compared to the 9-turn 
inductor.  Example current noise spectra through the superconducting transition for a prototype HF TES 
are shown in Figure~\ref{fig:MUXandTESnoise}.
While the TES designs for the two optical bands are different,
both exhibited these general trends in noise and stability, leading to
the choice of the $\sim$\SI{60}{\nano\henry} inductance as the target for every TES in the HF array.

Unlike single pixel measurements, TESes in the actual HF
array have additional inductance sourced by the extra wirebonds,
wiring, and cables connecting them to the interface chips.  In a cooldown
of a prototype HF array with a quarter of the readout fully populated,
we directly measured this additional inductance by comparing the time
domain response of the TES bias circuit to square pulses with the
TESes superconducting between the prototype array TESes and the single
pixel devices.  These measurements imply the extra wiring in the array
and array readout adds the equivalent inductance of an $\sim$8-turn
spiral inductor.  Estimates of the added series inductance of the
array and readout wiring of $23-$\SI{100}{\nano\henry} obtained using
the FastHenry inductance solver~\cite{Kamon94} are in excellent
agreement with $(44\pm28)$~\SI{}{\nano\henry}, the measured average and rms inductance for all
detectors in the partially integrated HF array prototype.  The expected added
inductance for each channel depends on the specific length of the
wiring connecting it from the array into the readout.
Because the array and readout wiring itself on average adds the targeted inductance 
needed to bandwidth limit the TES signals while maintaining TES 
stability, the HF interface chips have been bonded so they add no extra inductance.

The series inductance of this wiring bandwidth-limits the TES signals
and helps to prevent excess TES noise at frequencies above the
multiplexing rate from aliasing into our science band.  However, an
unavoidable source of noise in any TDM readout is out-of-band SQUID
and preamplifier noise above the multiplexing rate, which is aliased
into the signal band.  This noise source can be mitigated by
increasing the multiplexing rate, thereby decreasing the amount of
noise power above the Nyquist frequency which can be aliased into the
science band.  For ACTPol, $t_{dwell}$, the amount of time the
multiplexer spends on each row while switching and sampling, was
limited to $\gtrsim$\SI{2}{\micro\second}, limiting the multiplexing
rate to $\lesssim$\SI{15.15}{\kilo\hertz}.  Measurements of the total AdvACT
readout bandwidth shown in Figure~\ref{fig:MUXandTESnoise} indicate
that we still require $t_{dwell}\gtrsim$\SI{2}{\micro\second}, despite
the much higher demonstrated bandwidth
($t_{dwell}=$\SI{160}{\nano\second}) of the newer MUX
architecture~\cite{Doriese15}.  This is of particular concern for the
AdvACT HF array which has nearly twice as many channels as the ACTPol
arrays, limiting the maximum multiplexing rate to
\SI{7.8}{\kilo\hertz}.  While preliminary data indicates that readout
performance at \SI{7.8}{\kilo\hertz} will be sufficient to meet
\AdvACT's science goals, work is underway to increase the bandwidth of
the readout chain in order to enable higher multiplexing rates
for \AdvACT and future efforts.

\subsection{Multi-Channel Electronics}
\label{subsec:mce}

Originally developed for SCUBA2~\cite{Holland13}, the ambient
temperature Multi-Channel Electronics (MCE) used to multiplex the
\AdvACT arrays have been described extensively
elsewhere~\cite{Battistelli08,Battistelli08_2}, so we focus on the
changes that were required to enable the readout of the \AdvACT
arrays.  Each array is read out using one MCE control crate, and each
MCE crate contains nine cards ; one addressing card (AC), three bias
cards (BCs), four readout cards (RCs), and one clock card (CC).  The
AC contains 41 digital-to-analog converters (DACs) for switching rows
in the MUX, each BC contains 32 DACs for setting and switching SQUID
biases and feedbacks, the RCs implement the flux-locked loop (FLL)
used to maintain SQUID linearity and read out the TESes, and the CC
serves as an interface to and master for each MCE.  The BCs
additionally source low noise TES bias lines for the arrays.

Only minor firmware modifications and no hardware adjustments to the
MCE were required to extend the MUX factor from 33 in ACTPol to 64 for
\AdvACT.  This was possible because of the reduction in signaling
lines required by the new two-stage SQUID MUX architecture.  ACTPol's
three-stage SQUID MUX architecture required independent DACs to
specify biases and feedbacks for not only a first stage SQ1 and a
\SI{1}{\kelvin} SA, but also for an intermediate second-stage SQ2 on
every readout column.  While the two-stage architecture requires
additional DACs for switching the FASes on each row, it frees 64 DACs
by eliminating the SQ2, 32 of which are used in \AdvACT to bias SQ1s,
and 32 of which are used to switch additional rows.  This results in a
total of 33 rows switched from the AC as in ACTPol, and 32 additional
rows that can be switched using a BC.  While all 65 of
these AC and BC DACs are wired to FASes for the HF array, due to
firmware limitations only 64 can be multiplexed simultaneously.  One
special DSQ row in the multiplexer connected to channels that are
deliberately not connected to detectors will not be read out in normal
operation.

We have developed software transparently implementing this ``hybrid'' switching scheme across
the DACs of the AC and one BC.  While in principle TDM MUX factors as
high as 73 should be achievable using MCE hardware and this hybrid
switching scheme, going beyond 64 rows would require significant
firmware development to overcome hard-coded limits to the size of
memory blocks on the FPGAs of all the MCE cards.

In the old ACTPol MUX architecture, rows in the readout were
activated by switching the SQ1 bias ``on'' to a single fixed value for every
SQ1 in that row.  One of many benefits of the new architecture is that the SQ1 bias 
can now be varied at the multiplexing rate
ensuring that each SQ1 is individually and optimally 
biased for each row read out.  As in the older architecture, the series array feedback is also independently
multiplexed with the array, enabling each of the 2048 HF readout channels to be 
independently servoed to its own optimal SA operating point.

As in ACTPol, for AdvACT synchronization of data will be accomplished
using a ``sync box'' which provides timing signals to all MCEs and the
telescope housekeeping system, allowing bolometer and pointing
information to be precisely aligned~\cite{Thornton16}.  To allow the telescope to
operate with MCEs in both 64 and 33 row configurations for the
deployment of the HF array alongside two ACTPol 33-row arrays, the
sync box firmware has been altered to provide two, precisely
synchronized timing parameter streams.  The pointing information is
precisely aligned with one of the MCE streams, and for the other MCE
stream the pointing is obtained by interpolation.

\subsection{SQUID Tuning Procedure}
\label{subsec:tuning}
\begin{figure} [ht]
   \begin{center}
   \begin{tabular}{c}
     \includegraphics[width=16cm]{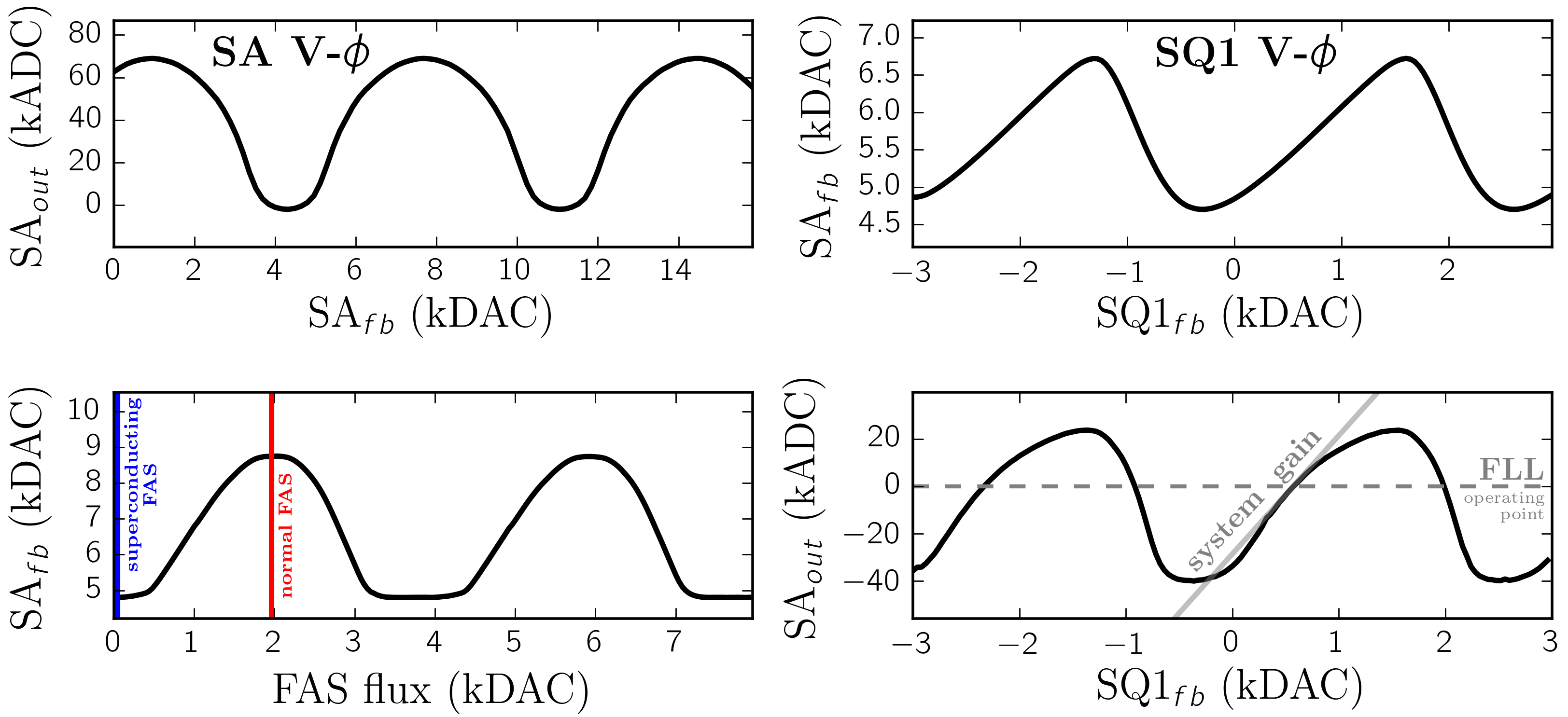}
   \end{tabular}
   \end{center}
   \caption[example] 
   { \label{fig:muxtuning} Plots of several stages of the MCE SQUID
     tuning procedure for the \AdvACT HF array, for one example
     channel.  \emph{Upper left:} SQUID Series Array (SA) \Vphi curve,
     generated by ramping current through the SA feedback coil and
     directly measuring the SA output voltage with the SA biased at
     $I_{c}^{max}$.
     \emph{Lower left:} The ``FAS servo'' tuning stage, generated by
     ramping current through the FAS flux coil for this channel and
     measuring the required SA feedback to keep the SA locked.  The
     flux at which the FAS is superconducting (normal) is determined
     from the minima (maxima) of this curve, and is used to
     switch channels off (on) while multiplexing the array.
     \emph{Upper right:} The ``SQ1 servo'' tuning stage, generated by
     ramping current through this channel's SQ1 feedback inductor and
     measuring the SA feedback required to keep the SA locked, providing
     a measurement of this SQ1's \Vphi.
     \emph{Lower right:} The ``SQ1 servo'' tuning stage, generated by
     ramping the SQ1 feedback for this channel with the SQUIDs
     unlocked, provides a sanity check of the detector servo operating
     point and a measure of the total readout system gain.
}
\end{figure} 

We have extended the automated software ``tuning'' procedure for
determining the best operating SQUID biases and feedbacks for the MCE
to the \AdvACT MUX architecture.  Over 10,000 parameters must be
chosen to multiplex the full HF array with the MCE, and full
automation of the procedure is critical to reduce deadtime due to
tuning during observations.  Tuning proceeds in stages and builds
heavily upon the extensive MCE software tools already developed for
ACTPol and other instruments~\cite{Battistelli08_2}.  Operating parameters
are determined from measuring the \Vphi curves of each SQUID in the
MUX, where $\phi$ is the flux coupled into each SQUID by ramping
current down its feedback coil.  Measured \Vphi from different
stages of a typical tuning are shown in Figure~\ref{fig:muxtuning},
along with some of the inferred operating parameters.

As for the old architecture, the \Vphi of the \SI{1}{\kelvin} SA is
measured first, to determine the optimal SA bias and operating point
for each readout column (this tuning stage is described in more detail in a
prior proceedings~\cite{Battistelli08}).  The SA operating point for each
column is an SA feedback bias at which the SA \Vphi is approximately 
linear and the measured SA voltage at that feedback is $V_{SA}$. 
During FLL operation the SA is held fixed at $V_{SA}$.
This is accomplished by adjusting the total flux coupled into the SA (either
directly via the SA feedback, or indirectly via the feedback of a lower coupled 
SQUID stage) to maintain the output voltage of the SA at $V_{SA}$.

Next, the FAS ``on/off'' bias conditions are found as follows.  The
SA feedback required to return the SA output voltage to $V_{SA}$
is determined as the FAS flux is ramped.  This SA feedback is proportional to 
the total current through the FAS and SQ1.
The FAS servo curves typically show ranges of FAS flux over
which the FAS response is flat and the FAS is superconducting, and
ranges of FAS flux over which the FAS exhibits a finite voltage drop,
and is resistive, voltage biasing its companion SQ1.  Because
the FASes are superconducting for zero applied flux, the FAS
superconducting (normal) flux is chosen as the first minimum (maximum)
of these FAS flux ramp curves.  These flux values are stored and the
MUX begins switching, sequentially setting each row to be readout to
its normal flux while pinning all other rows to their 
superconducting flux values.

Following the FAS servo stage is an ``SQ1 servo'' stage, identical 
to the former except that the SQ1 feedback is ramped instead of 
the FAS flux.  This is done while the FASes are being switched, 
resulting in per-channel SQ1 \Vphi measurements, which are
repeated over a range of pre-specified SQ1 biases.  The SQ1 servo
stage allows the determination of the optimal SQ1 bias that maximizes 
the peak-to-peak modulation of each channel's SQ1 \Vphi for
2048 SQ1 bias values in total.  In addition to an SQ1 bias for each channel,
this stage determines the SA feedback required to run each 
channel at the SA operating point.  The MUX then begins switching
the per-channel SA feedbacks and SQ1 biases to their optimal 
values in preparation for array readout.

The last stages of the tuning, besides initiating the TES FLL, maps the open
loop response of each channel by ramping the SQ1 feedback and
measuring the output SA voltage.  The data taken in this ``SQ1 ramp''
stage provide a sanity check for the configuration of every MUX
channel, and can be used to better optimize the tuning servos used in the FAS
servo and SQ1 servo tune stages.  A similar ``SQ1 Ramp TES''
diagnostic tuning stage measures the open loop response of the MUX to
ramping the voltage on all TES bias lines simultaneously and can be
used to optimize the FLL used to measure the TES signals.

\section{HF readout component characterization and screening}
\label{sec:characterization}
In total, 96 interface chips, 192 MUX chips, and
more than $20,000$ aluminum wirebonds are required to fully channelize the 2024
TESes in the HF array.  Below, we discuss the screening performed 
on all of the MUX and interface chips integrated into the \AdvACT HF array.
More details on the integration of the HF
array and its many components can be found in accompanying papers in
this proceedings~\cite{Li16,Ward16}.

\subsection{MUX chip screening data and analysis}
   \begin{figure} [ht]
   \begin{center}
   \begin{tabular}{c}
   \includegraphics[width=16cm]{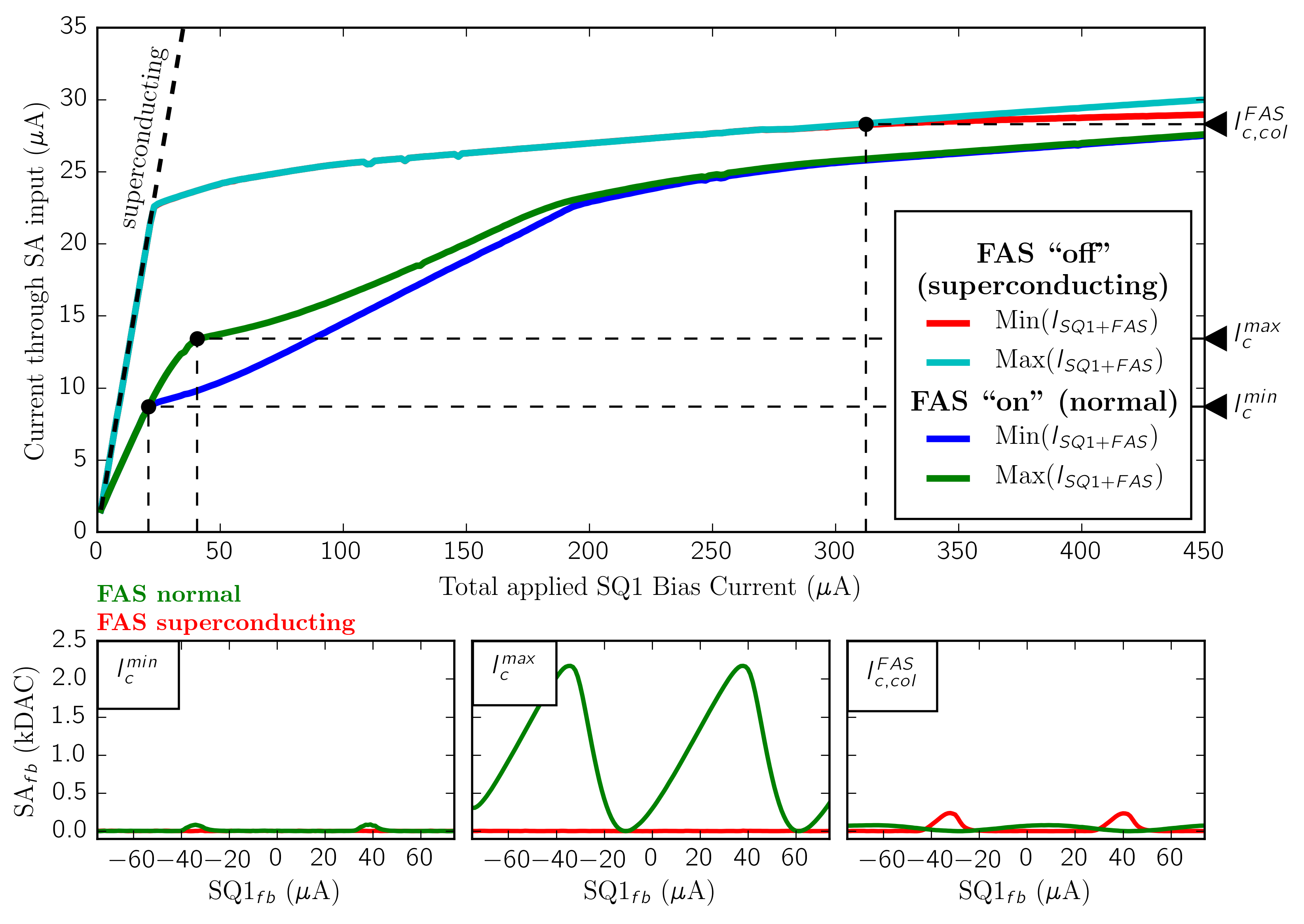}
   \end{tabular}
   \end{center}
   \caption[example] { \label{fig:muxtheory} \emph{Top:} 
     The minimum and maximum current through an example channel's SQ1
     and FAS, extremized over all values of applied SQ1
     feedback, measured from SQ1 servo curves as a function of SQ1
     bias.  Measurements are shown both with the FAS
     superconducting (cyan and red) and normal (dark green and dark
     blue).  With the FAS normal (channel on), the majority of the
     current is shunted through the channel's SQ1.  Arrows denote SQ1
     biases and currents corresponding to $I_c^{min}$, where the
     minimum and maximum current first diverge, and $I_c^{max}$ where
     the difference is extremized.  Also denoted is $I_{c,col}^{\rm{FAS}}$,
     the SQ1 bias at which the readout column containing this channel
     first exhibits persistence.  \emph{Bottom Left:} SQ1 servo curves
     for this channel with its FAS normal (green) and superconducting
     (red) at an SQ1 bias current just above $I_c^{min}$.
     \emph{Bottom Middle:} Same but with the SQ1 bias current at
     $I_c^{max}$.  At $I_c^{max}$ the channel's SQ1 is no longer
     superconducting for any applied SQ1 feedback.  \emph{Bottom
       Right:} Same but with the SQ1 bias current just above
     $I_{c,col}^{\rm{FAS}}$.  At these high SQ1 bias currents, at least one
     FAS on this channel's readout column is no longer superconducting
     for any applied flux and the MUX is no longer able to properly
     switch on this column.}
   \end{figure} 

For screening, multiplexing chips are assembled on an
independent test PCB in eleven columns of four chips each. The chips
are tuned in stages as described in \s{subsec:tuning} above, optimizing
the parameters required by each stage. Once optimized, a full tune is
taken at an SQ1 bias close to the expected optimal bias and the data
from each stage of the tuning are examined individually to flag
critical failures (open lines, unresponsive SQUIDs, etc.).  Next, full
tunes are taken on each readout column individually, slightly
incrementing the SQ1 bias current for each successive tune to map each
SQ1's \Vphi modulation as a function of SQ1 bias. At low SQ1 bias, no
modulation occurs.  There is an intermediate range of SQ1 biases where
the SQ1 \Vphi curves show modulation, but still have a superconducting
branch. At sufficiently high SQ1 bias, the superconducting branch
disappears.

The critical current $I_c^{min}$ is the smallest current applied to
the SQ1 bias for which modulation appears. The critical current
$I_c^{max}$ is the current for which the superconducting branch of the
SQ1 \Vphi curve first completely disappears.  Here, $I_c^{min}$ and
$I_c^{max}$ are defined as stated when the FAS is normal.  For this
first SQ1 bias sweep, the SQ1 bias is stepped between tunes by between
$73.5$ and \SI{294}{\nano\ampere} from zero up to $\sim2.5
I_c^{max}$, typically corresponding to steps of 0.5\% and 2\% of
$I_c^{max}$, respectively (later screens were done at higher
resolution in SQ1 bias).  $I_{\rm{FAS}}$ is the amount of current applied
to the FAS flux input which switches a particular channel of the MUX
on.  The goals of the screening are to ascertain if a MUX chip is
functioning normally and to determine these three parameters so that we can
optimize which MUX chips will be integrated together into readout columns.

The process of taking tunes while ramping the SQ1 bias is then
repeated, but this time with coarser steps and a much higher maximal
value of SQ1 bias ($\gg I_c^{max}$), and with all FASes
switched off.  The purpose of this ramp is to check for
``persistence'' near or below $I_c^{max}$: a row may remain on for all
values of
$I_{\rm{FAS}}$, interfering with the readout of every other row on a
column.  Although this is a failure which affects the entire readout
column, it is usually possible to identify the offending channel.
This is cause to mark a MUX chip defective, and is typically diagnosed
by observing an unusual FAS servo tuning curve.  If the chip with the
persistent channel can be identified, the other chips on the column
must be retested in a subsequent cryogenic screening. At very high SQ1 biases ($\gg I_c^{max}$), even
channels which do not exhibit persistence at or near $I_c^{max}$ will
become persistent.  We require that this critical current,
$I_{c,col}^{\rm{FAS}}$, be $\geq 2 I_c^{max}$ for the chips on a
screening column to be included in the HF array.  This additional
selection was imposed partway through screening and only approximately
one half of the MUX chips integrated into the HF array satisfied this
inequality, although every chip integrated into the array had
$I_{c,col}^{\rm{FAS}}/I_c^{max}>1$ (only lower limits on
$I_{c,col}^{\rm{FAS}}/I_c^{max}$ were obtained for these chips).
Figure~\ref{fig:muxtheory} illustrates how $I_c^{min}$, $I_c^{max}$
and $I_{c,col}^{\rm{FAS}}$ are determined from the aforementioned SQUID
current and voltage sweeps.

Other parameters of interest for each MUX channel are determined from the
first sequence of tunes versus SQ1 bias with the FASes switching
normally.  In particular, $I_{\rm{FAS}}$ for each FAS is determined from
the values used in the tune nearest $I_c^{max}$, since in operation
each SQ1 will be biased near its $I_c^{max}$ value.

The algorithms for determining $I_c^{min}$ and $I_c^{max}$ from these
data are very similar. First, the minimum (for $I_c^{max}$) or maximum
(for $I_c^{min}$) of the SQ1 servo curves is determined for each SQ1
bias.  Physically, the ordinate of the SQ1 servo curves corresponds to
the current through the SQ1 (very little current is sourced by a
channel's FAS when it is normal) and is calibrated from SA feedback in
digital counts to physical current using the measured DC impedance of
the SA and SQ1 cold circuits and the ratio of the mutual inductances
of the SA feedback and input coils.  The current through the SQ1
varies approximately linearly with SQ1 bias current, if the impedance
of the SQ1 loop relative to its \SI{1}{\ohm} shunt resistor is
constant.  Discontinuities occur when the impedance in the SQ1 loop
changes, as when modulation first appears at $I_c^{min}$ or when the
SQ1 superconducting branch first disappears at $I_c^{max}$, as shown
in Figure~\ref{fig:muxtheory}.  We determine $I_c^{min}$ and
$I_c^{max}$ by identifying the currents through the SQ1 at which these
curves exhibit these discontinuities.  The actual values of
$I_c^{min}$ and $I_c^{max}$ are determined by fitting these curves
below and above their corresponding discontinuities to lines and
computing the SQ1 current at these lines' intersection points.

\subsection{HF MUX chip screening results}

After screening, only chips passing the selection criteria are
considered for array integration. The majority of chips fail the
screening due to wiring errors, dead channels, channels with unusual
looking SQUID curves (either FAS or SQ1), a value of $I_{c,col}^{\rm{FAS}}$
less than $2 I_c^{max}$ (for chips screened later), a range of
$I_c^{max}$ values among its eleven rows of more than 25\% from the
chip's mean $I_c^{max}$, or damage sustained in handling or
fabrication.  All of the channels on chips that pass these criteria
tend to have similar values for $I_c^{min}$ and $I_c^{max}$, but
between these chips these values may vary. 
In particular, the mean and rms $I_c^{max}$ for chips from the three
wafers screened for the HF array, W5, W6 and W7, were $(9.9\pm0.6)$~\SI{}{\micro\ampere}, $(13.5\pm0.6)$~\SI{}{\micro\ampere},
and $(12.8\pm0.7)$~\SI{}{\micro\ampere}, respectively.  For passing
chips, $I_{\rm{FAS}}$ values were nearly identical, with a mean and rms of
$(240\pm 14)$~\SI{}{\micro\ampere}.
It is critical for the $I_{\rm{FAS}}$ of
FASes on the same row to be as similar as possible for the channels to
switch properly.

\begin{figure} [ht]
\centering
\begin{tabular}{@{}c@{}}
\includegraphics[width=16cm]{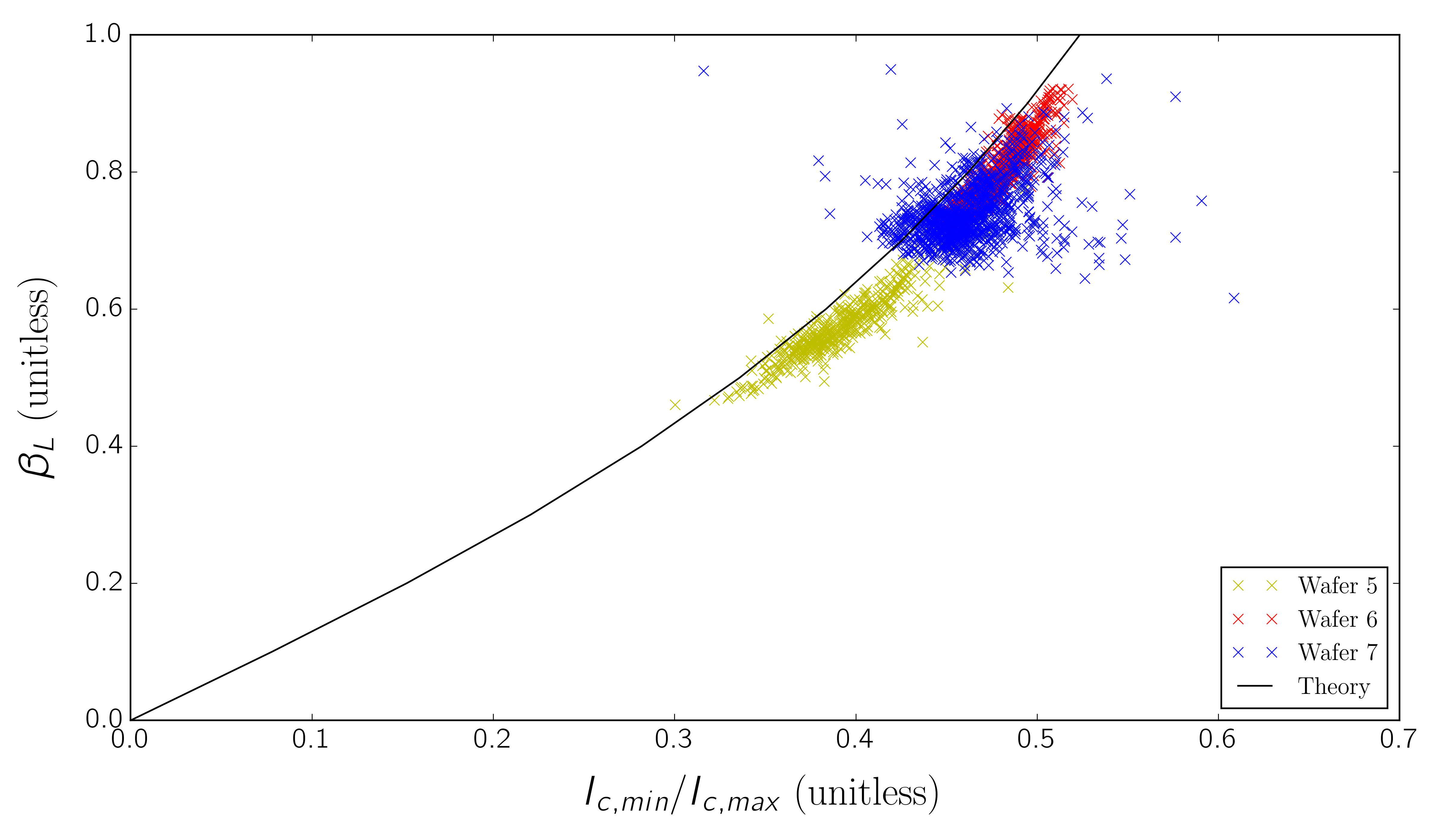}
\end{tabular}

\caption{The measured SQUID screening parameter $\beta_L$ vs
  $I_c^{min} / I_c^{max}$ for each channel, organized by wafer for MUX
  chips which passed screening for the AdvACT HF array.  Channels
  which operate as designed are expected to fall along the theory
  curve, which is a numerical solution to the transcendental equation
  relating $\beta_L$ to the ratio of $I_c^{min} /
  I_c^{max}$ from Reinhold and Koelle~\cite{Clarke05}.
}\label{fig:betaL}
\end{figure}

Reinhold and Koelle~\cite{Clarke05} give a
transcendental equation relating the ratio of $I_c^{min}/I_c^{max}$
to the SQUID screening parameter $\beta_L\equiv 2LI_{0}/\Phi_{0}$, where $L$
is the SQUID loop inductance, $I_{0}$ is the critical current of each
SQUID Josephson junction, and $\Phi_{0}$ is the magnetic flux
quantum.  A cross-check to validate our screening
procedure is to plot $I_c^{min}/I_c^{max}$ against the parameter
$\beta_L$ and compare it to a numerical solution of the theoretical
prediction. Such a plot for each channel on chips which passed the HF
screening can be seen in Figure~\ref{fig:betaL}.  To compute
$\beta_{L}$, we assume that $L=$\SI{120}{\pico\henry}, the nominal
design SQ1 loop inductance.

Chips with similar $I_c^{max}$ values are grouped together on a column
for integration. Although the SQ1 bias values are multiplexed, this
helps to avoid creating columns where one chip's $I_c^{max}$ is
greater than another chip's $I_{c,col}^{\rm{FAS}}$. Columns for the array
were chosen by first sorting screened chips by their $I_c^{max}$
values and then grouping consecutive chips in the sort into 6-chip 
readout columns.

\subsection{HF interface chip screening}
A total of 130 interface chips were screened for the HF array, with only
96 chips selected for final integration.  Chips were warm probed for
shorts between adjacent channels, and the summed resistance of all 22
shunt resistors on each interface chip was measured cryogenically
using a four-lead configuration in a dilution refrigerator at
\SI{100}{\milli\kelvin}.  Common reasons for an interface chip to fail
screening included unintentional shorts between adjacent channels due
to fabrication errors and abnormal cryogenic resistance.  The target
per-channel shunt resistance $R_{sh}$ was \SI{180}{\micro\ohm}.  The
interface chips in the HF array came from four 3$\inch$ wafers of 40
interface chips apiece fabricated at NIST/Boulder.  The average
$R_{sh}$ over three wafers was \SI{203}{\micro\ohm} with no more than
5\% deviation for any measured chip (although one anomalous fourth wafer, W2,
had an average $R_{sh}$ of \SI{156}{\micro\ohm}).  Chips from all four
wafers were integrated into the HF without mixing
chips from W2 with chips from other wafers on the same TES bias line.
Chips were further grouped on each bias line by their measured
\SI{100}{\milli\kelvin} resistances, requiring less than a $3$\%
deviation in total cryogenic resistance for chips on the same TES bias
line.

\section{Summary}
\label{sec:summary}

The readout components for the first \AdvACT array have been screened,
and integration is complete.  We have described the design and
performance of the readout electronics for the array, and this readout
has been used successfully to characterize the \AdvACT HF array, which
will be deployed in mid-2016 and observe the CMB at
\SI{150}{\giga\hertz} and \SI{230}{\giga\hertz}.

The readout techniques developed for \AdvACT and described in this proceedings
enables the readout of significantly larger arrays of TES bolometers
than was previously possible using an MCE.  While other multiplexing
schemes employing alternate technologies are currently under development
and could yield MUX factors an order of magnitude larger
than the 64 MUX factor demonstrated here,~\cite{Day03,Irwin04,Kher16}
current efforts that rely on low-frequency TDM to multiplex large
arrays of TES bolometers stand to benefit substantially from larger
MUX factors.

\acknowledgments
SWH thanks Jeffrey Filippini and the BICEP3 collaboration for early
assistance.  
SWH and SPH thank Ti-Yen Lan for voluntarily couriering
MUX test boards between Princeton and Cornell.
This work was supported by the U.S. National Science
Foundation through award 1440226. The development of multichroic
detectors was supported by NASA grants NNX13AE56G and NNX14AB58G. The
work of \NSTRFGradStudents{} was supported by NASA Space Technology
Research Fellowship awards.

% References
\bibliography{advact_spie16_swh}
\bibliographystyle{spiebib} % makes bibtex use spiebib.bst
\end{document}